\documentclass[%
 reprint,
 amsmath,amssymb,amscd
 aps,
]{revtex4-1}

\usepackage{graphicx}
\usepackage{dcolumn}
\usepackage{bm}

\begin{document}

\preprint{APS/123-QED}

\title{Finding new signature effects on galactic dynamics \\to constrain Bose-Einstein-condensed cold dark matter}

\author{Tanja Rindler-Daller}
\thanks{Invited plenary speaker at the IV International Meeting on \\Gravitation and Cosmology, Guadalajara, MX, May 21-25, 2012}
\author{Paul R. Shapiro}%
\affiliation{%
 Department of Astronomy and Texas Cosmology Center, The University
 of Texas at Austin, 2515 Speedway C1400, Austin, TX 78712, USA
}%

\date{\today}

\begin{abstract}
 If cosmological cold dark matter (CDM) consists of light enough
 bosonic particles that their phase-space density exceeds unity,
 they will comprise a Bose-Einstein condensate (BEC). The nature of
 this BEC-CDM as a quantum fluid may then distinguish it dynamically
 from the standard form of CDM involving a collisionless gas of
 non-relativistic particles that interact purely gravitationally. We
 summarize some of the dynamical properties of BEC-CDM that may lead
 to observable signatures in galactic halos and present some of the
 bounds on particle mass and self-interaction coupling strength that
 result from a comparison with observed galaxies.
\begin{description}
\item[PACS numbers]
95.35.+d; 95.30.Cq; 98.62.Gq; 03.75.Nt

\end{description}
\end{abstract}

\maketitle


\section{\label{sec:level1}Introduction}

Astronomical observations suggest the presence of non-baryonic,
non-relativistic ('cold') dark matter (DM), comprising around 23 per
cent of the energy density in the Universe. The particle nature of
dark matter remains elusive, however, despite ongoing efforts to
detect it directly in search experiments or indirectly via imprints
on astrophysical observations. In large N-body simulations of
structure formation, cold dark matter (CDM) has been modeled as a
collisionless gas, which only interacts gravitationally. We refer to
this as standard CDM. Despite the successes in reproducing the
large-scale structure, as well as reproducing flat rotation curves
at the outskirts of galaxies, standard CDM seems to be in conflict
with observations of galactic small-scale properties. These are
notably the overabundance of subhalos around big hosts of Milky-Way
size and beyond, and the failure to reproduce flat cores in the
centers of dark-matter dominated dwarf and LSB galaxies. Both
features apparently contradict astronomical observations, and have
been the subject of active research in the past decade. One approach
has been to study the combined effects of the collisionless CDM
component and the dissipative baryonic component, to see if the
dissipative hydrodynamics of the latter can affect a cure. Another
possible solution has been to go beyond the simplifying assumption
that the DM particles are cold and/or collisionless. As a result,
there has been a recent revival of investigations of structure
formation with non-standard dark matter candidates, like warm dark
matter e.g. Schneider et al.\cite{schneider}, self-interacting
fermionic dark matter (SIDM) e.g. Ahn \& Shapiro \cite{AS}, Koda \& Shapiro
\cite{KS} and Bose-Einstein-condensed dark matter (or scalar-field
dark matter), e.g. Woo \& Chiueh \cite{woo} and Su\'arez \& Matos
\cite{matos}, all proposals of which are able to suppress the
formation of self-bound structures below a certain scale, as well as
to flatten central profiles, depending on the respective particle
parameters.

In this paper, we will address Bose-Einstein-condensed DM (BEC-CDM).
We are particularly interested in a class of models describing
self-interacting bosonic dark matter, the particles of which are so
light that they collectively occupy their ground state below a
certain temperature, forming a Bose-Einstein condensate (BEC) in the
early Universe. Then, this state can be well described by a scalar
field, the so-called wavefunction of the condensate. We will assume
that the wavefunction respects a $U(1)$-symmetry, such that the
number of particles is conserved. There is thus no self-annihilation
of DM in this scenario, in contrast to models considered for
instance by Tkachev \cite{tkachev}. Recent advances in particle
theory predict the generic existence of bosons as light as or much
lighter than the QCD axion - the 'classic' bosonic DM candidate,
values ranging from about $10^{-33}$ to $\gtrsim 10^{-5}$ eV/c$^2$,
which can serve as the CDM in the Universe (see e.g. G\"unther \&
Zhuk \cite{GZ}, Carroll \cite{carroll}, Arvanitaki et al.
\cite{axiverse}). The theoretical description of these very light
bosons in terms of scalar fields leads to halo dynamics which can be
understood as the solution of nonlinear wave equations (e.g.
Alcubierre et al.\cite{alcubierre}, Chavanis \cite{chavanis1}, Matos
\& Ure\~na-L\'opez \cite{MU}, Sin \cite{sin}, Ure\~na-L\'opez \&
Guzm\'an \cite{urena}), in contrast to the $N$-body dynamics of
standard CDM. As a matter of fact, much less investigation has been
pursued in the literature so far, in particular with respect to the
nonlinear stages of structure formation for this type of DM.
Therefore, we are in need of a better understanding of how much of
the parameter space of this form of dark matter is able to reproduce
the successes, while resolving the failures, of standard CDM.

In what follows, we will summarize some of the recent work in this
field, highlighting some of our own, with apologies for the fact
that length limitations prevent us from attempting a more
comprehensive review. In Section \ref{sec:level2}, we will describe
the basic equations which govern BEC-CDM dynamics, showing how its
quantum nature leads to fluid behavior. We will then distinguish two
regimes according to the strength of the particle self-interaction,
and state how the characteristic size and mass of structures that
form gravitationally from this form of DM are related to the
particle mass $m$ and self-interaction coupling strength $g$. We
also present the virial theorem for isolated BEC-CDM halos. In
Section \ref{sec:level3}, we summarize how the equilibrium structure
of BEC-CDM halos can distinguish them from halos in standard CDM,
for halos with and without rotation. We show that halos in the
BEC-CDM model typically rotate with enough angular momentum that, if
they are of the type supported against gravitational collapse by
self-interaction pressure and rotation, quantum vortices are likely
to form, which can affect halo density profiles. We also summarize a
few examples in which BEC-CDM can be distinguished by its effect on
baryonic structures. In Section \ref{sec:level4}, we summarize the
bounds on particle mass and coupling strength that follow from the
requirement that halos or halo cores not exceed some characteristic
size, if halos or their cores are treated as isolated, equilibrium
structures. Finally, in Section \ref{sec:level5}, we argue that, for
BEC-CDM to reproduce the full range of structure scales found in our
Universe (as in standard CDM), we must go beyond the modeling of
individual objects in static equilibria to account for continuous
mass infall and its dynamical consequences.

\section{\label{sec:level2}Quantum-coherent dark matter under Newtonian gravity}

\subsection{\label{sec:leve21}Fundamental properties}

The dynamical description of dark matter within galactic halos
usually involves small gravitational and velocity fields. With
regard to BEC-CDM, it has proved advantageous to consider the
non-relativistic Schr\"odinger-Poisson, or Gross-Pitaevskii-Poisson
(GPP) system of equations of motion for the dark matter BEC
wavefunction $\psi$, as follows
\begin{equation} \label{gp}
 i\hbar \frac{\partial \psi}{\partial t} = -\frac{\hbar^2}{2m}\Delta \psi + m\Phi \psi +
 g|\psi|^2\psi,
 \end{equation}
 \begin{equation} \label{poisson}
  \Delta \Phi = 4\pi G m |\psi|^2.
   \end{equation}
The terms on the rhs in (\ref{gp}), which govern the evolution, are
due to the quantum-kinetic energy, the gravitational potential
$\Phi$, and the self-interaction of identical bosons. The latter has
been included in the usual way in terms of an effective interaction
potential $g|\psi|^4/2$ with coupling constant (or self-interaction
strength) $g$. The possibly complicated particle interactions are
simplified this way in the low-energy limit of a dilute gas:
disregarding higher than 2-body interactions, the cross section for
elastic scattering of indistinguishable bosons becomes constant in
the low-energy limit,
\begin{equation} \label{sigma}
  \sigma = 8\pi a_s^2
  \end{equation}
with the s-wave scattering length $a_s$. The coupling constant of
the effective interaction is then given by
\begin{equation} \label{coupling}
  g = 4\pi \hbar^2 a_s/m.
   \end{equation}
  We shall note that the above GP equation is strictly
valid only for dilute systems, which means that $a_s$ must be much
smaller than the mean interparticle distance, i.e. $a_s \ll
n^{-1/3}$. Also, we are restricted to $g \geq 0$ ($a_s \geq 0$),
because models described by (\ref{gp}) with negative
self-interaction coupling $g$ are not able to provide stable
structures, since the resulting negative pressure works in favor of
gravity.
   If we assume that all of
the DM is in the condensed state, the number and mass density of DM
in the halo is given by $n(\mathbf{r}) = |\psi|^2(\mathbf{r})$ and
$\rho(\mathbf{r}) = m n(\mathbf{r})$, respectively.

The equations can be written in fluid-like form by inserting into
(\ref{gp}) the polar decomposition of the wavefunction (applied
early by Madelung \cite{madelung} to the free-particle Schr\"odinger
equation),
  \begin{equation} \label{polar}
\psi(\mathbf{r},t) = |\psi|(\mathbf{r},t)e^{iS(\mathbf{r},t)} =
\sqrt{\frac{\rho(\mathbf{r},t)}{m}}e^{iS(\mathbf{r},t)},
 \end{equation}
 resulting in the momentum and continuity equations,
  \begin{equation} \label{fluid}
     \rho \frac{\partial \mathbf{v}}{\partial t} + \rho (\mathbf{v} \cdot \nabla)\mathbf{v} = -\rho \nabla
     Q
    - \rho \nabla \Phi - \nabla P_{SI},
     \end{equation}
 \begin{equation} \label{hd3}
\frac{\partial \rho}{\partial t} + \nabla \cdot (\rho \mathbf{v}) =
0,
 \end{equation}
   with the bulk velocity $\mathbf{v} = \hbar \nabla S/m$.
The gradient of
  \begin{equation} \label{qpot}
   Q = -\hbar^2 \Delta \sqrt{\rho}/(2m^2 \sqrt{\rho})
    \end{equation}
  gives rise to what is often called 'quantum
 pressure', an additional force on the rhs of equ.(\ref{fluid}), which basically stems from the quantum-mechanical
 uncertainty principle. The particle self-interaction, on the other hand, gives rise to a pressure
 of
 polytropic form
  \begin{equation} \label{selfpressure}
 P_{SI} =  g \rho^2/(2m^2).
  \end{equation}
The quantum-kinetic term provides an important characteristic length
scale, as follows: with $\Delta$ having a dimension of L$^{-2}$ and
changing to the momentum representation,
 one can easily see that the characteristic length is essentially nothing but the de Broglie length of the bosons
  \begin{equation} \label{deB}
 L \sim \lambda_{deB} = h/p = h/(mv).
  \end{equation}
Since we will consider virialized, isolated objects, it makes sense
to use the corresponding virial velocity in this expression for
$\lambda_{deB}$, assuming that the particles stay in their
Bose-Einstein-condensed state after virialization, which is the case
for the model parameters we are going to encounter in this paper
(see also Section \ref{sec:level4}). BEC-CDM without
self-interaction, $P_{SI} = 0$, was termed 'Fuzzy Dark Matter' by
Hu, Barkana \& Gruzinov \cite{hu}. In this review, we will call it
BEC-CDM of TYPE I. In this regime, it is the quantum-kinetic term
via (\ref{deB}) which determines the equilibrium size of
self-gravitating objects - in order for $\lambda_{deB}$ not to
exceed a certain galactic length scale, the particle mass $m$ must
be sufficiently large. Table \ref{tab:table1} contains different
typical halo sizes and the corresponding lower limits on $m$.
\begin{table}[tb]
\caption{\label{tab:table1}
 Lower bound on the boson mass, provided by the 'Fuzzy Dark Matter' regime, or TYPE I BEC-CDM, for different cosmological structures}
\begin{ruledtabular}
\begin{tabular}{llll}
&\multicolumn{1}{c}{\textrm{halo
mass}}&\multicolumn{1}{c}{\textrm{size}}&
\multicolumn{1}{c}{\textrm{boson mass}}\\
 & [$M_{\odot}$] & \textrm{[kpc]} & \textrm{[eV]}\\
 \hline
 \colrule
\textrm{Milky Way (MW)} & $10^{12}$ & $100$ & $1.066\cdot 10^{-25}$\\
\textrm{Dwarf Galaxy (DG)} & $10^{10}$ & $10$ & $3.371 \cdot 10^{-24}$\\
\textrm{Dwarf Spheroidal (dSph)} & $10^8$ & $1$ & $1.066\cdot 10^{-22}$\\
\textrm{Minihalo (MH)} & $10^6$ & $0.1$ & $3.371\cdot 10^{-21}$\\
\end{tabular}
\end{ruledtabular}
\end{table}
However, it has been observed in the previous literature, e.g.
Colpi, Shapiro \& Wasserman \cite{CSW} and Lee \& Lim \cite{LL}, and
we confirm it as well, see equ.(\ref{relate}) and Section
\ref{sec:level3}, that a larger mass $m$ than inferred from
(\ref{deB}) can result in stable structures of the same given size,
if self-interaction is included. In fact, considering the case in
which the last term in (\ref{fluid}) supports the system against
gravitational collapse, while $Q = 0$, we arrive at the opposite
regime to TYPE I, which we call the Thomas-Fermi regime of BEC-CDM,
or TYPE II for short. We note that this regime goes under many
names; Goodman \cite{goodman} calls it 'repulsive dark matter'
(RDM), Peebles \cite{peebles} speaks of 'fluid dark matter'. We also
note that both regimes have already been considered for related
models by Khlopov, Malomed \& Zeldovich \cite{khlopov}, in studying
gravitational instabilities of a primordially produced scalar field.

In the following, we will make use of convenient units, defined as
in Rindler-Daller \& Shapiro (RDS) \cite{RS}:
 \begin{displaymath}
  m_H \equiv \frac{\hbar}{R^2(\pi G \bar \rho)^{1/2}} =
  \frac{2\hbar}{\sqrt{3G}}(RM)^{-1/2} =
   \end{displaymath}
 \begin{equation} \label{mfidu}
    = 1.066 \cdot 10^{-22}\left(\frac{R}{1 ~\rm{kpc}}\right)^{-1/2}\left(\frac{M}{10^{8}~ M_{\odot}}\right)^{-1/2}
  \rm{eV},
   \end{equation}
    and
 \begin{displaymath}
  g_H \equiv \hbar^2/(2\bar \rho R^2) = 2\pi \hbar^2 R/(3M) =
   \end{displaymath}
     \begin{equation} \label{gfidu}
   = 2.252 \cdot 10^{-62} \left(\frac{R}{1~
   \rm{kpc}}\right)\left(\frac{M}{10^{8}~M_{\odot}}\right)^{-1}
   \rm{eV} ~\rm{cm}^3,
 \end{equation}
 with $c=1$. We described the meaning and significance of those parameters
at length in RDS \cite{RS}. It shall be sufficient to re-iterate
here that $m_H$ is the characteristic mass of a non-interacting
particle whose de Broglie wavelength is comparable to the size of a
given halo, see Table \ref{tab:table1}. It is thus the smallest
particle mass possible in order for quantum pressure to be solely
responsible for holding that halo up against gravitational collapse.
On the other hand, if there are density variations in the BEC fluid
of scale length $R$, then $g_H$ is the coupling strength for which
the quantum and self-interaction pressure force terms are equal.

It turns out that the TYPE II regime is a good approximation as long
as $g/g_H \gg 2$ is fulfilled, as we demonstrated in \cite{RS}: to
determine whether a BEC-CDM halo of a given size $R$ is of TYPE I or
TYPE II, we have to compare the quantum pressure and
self-interaction pressure terms in (\ref{fluid}) to each other,
 \begin{displaymath}
  |-\rho \nabla Q| / |-\nabla P_{SI}| \sim \hbar^2/(g\rho R^2)
  \sim
  2 g_H/g \ll 1,
  \end{displaymath}
 from which the claim follows. We have shown in RDS \cite{RS}, equ.(46), that the radius of a
spherical halo is then related to the de-Broglie length of the boson
according to
  \begin{equation} \label{relate}
   R_0 =
   \frac{\sqrt{3}\pi^{1/4}}{12}\left(\frac{g}{g_H}\right)^{1/2}\lambda_{deB}.
    \end{equation}
Hence, since $g/g_H \gg 2$, $R_0 \gg \lambda_{deB}$ for TYPE II
BEC-CDM halos.

\subsection{\label{sec:leve22}Stationary systems and virial equilibrium}

In the context of BEC-CDM in the GPP framework,
equ.(\ref{gp})-(\ref{poisson}), stationary self-gravitating halos
can be described by wavefunctions of the form
\begin{equation}
\psi(\mathbf{r},t) = \psi_s(\mathbf{r})e^{-i\mu t/\hbar},
 \end{equation}
where the conservation of particle number fixes $\mu$, the GP
chemical potential. While $\psi$ evolves harmonically in time, the
mass density $\rho = m|\psi_s|^2$ and, hence, the gravitational
potential $\Phi$ are time-independent. Inserting this $\psi$ into
(\ref{gp}) results in the time-independent GP equation with
eigenvalues $\mu$,
 \begin{equation} \label{stat}
 \left(-\frac{\hbar^2}{2m}\Delta + g|\psi_s|^2 + m \Phi\right)\psi_s = \mu
 \psi_s.
  \end{equation}
 The time-independent part $\psi_s(\mathbf{r})$ itself can be decomposed
 as usual,
   \begin{equation} \label{statdec}
   \psi_s(\mathbf{r}) = |\psi_s|(\mathbf{r})e^{iS_s(\mathbf{r})}
    \end{equation}
   with both amplitude and phase depending here on position only. We
   will omit the subscript 's' in the forthcoming analysis.
 Systems obeying (\ref{stat}) can
 be studied via the corresponding GP energy functional given by
 \begin{equation} \label{energie}
   \mathcal{E}[\psi] = \int_V \left[\frac{\hbar^2}{2m}
 |\nabla \psi|^2 + \frac{m}{2}\Phi |\psi|^2 +
 \frac{g}{2}|\psi|^4\right]d^3\mathbf{r}.
 \end{equation}
Inserting (\ref{statdec}) into (\ref{energie}), the total energy
  can be written as
  \begin{equation} \label{sumenerg}
   E = K + W + U_{SI},
    \end{equation}
 with the total kinetic energy term
  \begin{displaymath}
   K \equiv \int_V \frac{\hbar^2}{2m}|\nabla \psi|^2 d^3\mathbf{r} =
    \end{displaymath}
   \begin{equation} \label{kname}
    = \int_V
   \frac{\hbar^2}{2m^2}(\nabla \sqrt{\rho})^2d^3\mathbf{r} + \int_V
   \frac{\rho}{2}\mathbf{v}^2d^3\mathbf{r} \equiv K_Q + T.
    \end{equation}
 $K_Q$ accounts for the quantum-kinetic energy and $T$ for the
 bulk kinetic energy of the body, which comes in the form of rotation or internal motion. $K_Q$ has no
 classical counterpart, and is absent in the classical figures of equilibrium
studied in the previous literature.
 Also, $K_Q$ is neglected in the TYPE II regime.
  The other terms in (\ref{sumenerg}) are the gravitational
 potential energy
  \begin{equation} \label{wname}
   W \equiv \int_V \frac{\rho}{2}\Phi d^3\mathbf{r}
    \end{equation}
 and the internal energy
  \begin{equation} \label{internal}
   U_{SI} \equiv \int_V \frac{g}{2m^2}\rho^2 d^3\mathbf{r},
    \end{equation}
 which is determined by the particle interactions, and which we have defined essentially as
 $U_{SI} = \int P_{SI} dV$ with
 $P_{SI}$ in (\ref{selfpressure}). The origin of $U_{SI}$ is due to the repulsive 2-body
  elastic scattering of identical bosons, equ. (\ref{sigma}).
  The above energy
 contributions enter the scalar virial theorem of an \textit{isolated} (possibly rotating) BEC halo under self-gravity, which reads as
  \begin{equation} \label{virial}
 2K + W + 3U_{SI} = 0.
  \end{equation}
As in classical gas dynamics, (\ref{virial}) can be derived by
multiplying the equations of motion in fluid form,
equ.(\ref{fluid}), by $\mathbf{r}$ and integrating the resulting
equation over volumes which enclose the system of interest. For an
isolated body, a derivation involving a scaling argument was
presented by Wang \cite{wang}.

\section{\label{sec:level3}Signature effects of BEC-CDM on halos and halo cores}

\subsection{\label{sec:leve31}Sizes and density profiles}

The equilibrium density profiles of self-gravitating BEC-CDM halos
are solutions of (\ref{gp}) or (\ref{fluid})-(\ref{selfpressure})
respectively, along with (\ref{poisson}). As an important result,
they are universal in shape. Furthermore, BEC-CDM halos have a
finite central density. In fact, this last feature has been one of
the motivations in the previous literature to consider this form of
DM as a solution to the cusp-core problem of dark matter dominated
dwarf galaxies (DG) and dwarf-spheroidal (dSph) galaxies.

In the case of TYPE I, the density profile can only be determined
numerically: it falls off as $r^{-4}$ for large $r$, but has no
compact support. The radius which includes 99 per cent of the mass
reads as
 \begin{equation} \label{fuzzysize}
   R_{99} = 9.9 \hbar^2/(G M m^2)
    \end{equation}
 (see Membrado, Pacheco \& Sa\~nudo \cite{MPS} for more details).
For TYPE II, on the other hand, the equation of state reduces to an
($n=1$)-polytrope, with $P_{SI}$ in (\ref{selfpressure}) and $Q =
0$, having the well-known spherical density profile
 \begin{equation} \label{tfprofile}
   \rho^S(r) = \rho_c^S sinc (\sqrt{4\pi G m^2/g}~ r),
    \end{equation}
with $sinc(x) \equiv sin(x)/x$ and the central density $\rho_c^S$.
The corresponding halo radius is then given by
 \begin{equation} \label{onesphere}
   R_0 = \pi
   \sqrt{\frac{g}{4\pi G m^2}},
    \end{equation}
   see e.g. Goodman \cite{goodman} for more details.

In both regimes, TYPE I and TYPE II, the halo profile and size are
determined by the DM particle parameters. Of course, this is also
true for the intermediate regime: Chavanis \& Delfini \cite{CD}
calculate numerical solutions for the mass-radius relationship, $R =
R(M)$, which interpolate between TYPE I and II. However, for any
given particle model, neither (\ref{fuzzysize}) and
(\ref{onesphere}) nor the result of \cite{CD} are able to re-produce
the fact that $R$ must increase with $M$ as we know from
astronomical observations. The successful fitting of galaxy data
using the associated rotation curves of TYPE I and TYPE II BEC-CDM
halos by Arbey, Lesgourgues \& Salati \cite{arbey} and B\"ohmer \&
Harko \cite{BH} must thus be judged with this caveat in mind. It
implies that it is necessary to go beyond the description of
non-rotating halos in virial equilibrium, composed of a pure BEC-CDM
fluid, if this DM model is to describe galactic structures
successfully.

\subsection{\label{sec:leve32}Rotation and shape}

In RDS \cite{RS}, we studied the properties of BEC-CDM halos in the
TYPE II case, once rotation is taken into account. It is generally
believed that tidal torques caused by large-scale structure give a
halo most of its angular momentum in the early phases of halo
collapse. This picture has been confirmed by cosmological N-body
simulations of the standard CDM universe, which show that halos form
with a net angular momentum such that the dimensionless ratio, the
so-called spin parameter,
 \begin{equation} \label{lam}
  \lambda = \frac{L |E|^{1/2}}{GM^{5/2}},
   \end{equation}
    where $L$ is the total angular momentum and $E$ the total
    energy of the halo,
has typical values in the range of $[0.01, 0.1]$ with a median value
$\simeq 0.05$ (see e.g. Barnes \& Efstathiou \cite{BE}). The degree
of rotational support is thus very small for the CDM halos which
surround galaxies. We will be interested in the case where the BEC
nature of DM affects small-scale structure and the internal dynamics
of halos, while large-scale structure formation shall follow the
$\Lambda$CDM model to a great extent. Therefore, we adopt the above
range of spin parameters for BEC-CDM halos, too.

We describe the effect of rotation on the structure of BEC-CDM halos
by two approximations which are based upon the classic models of
rotating figures of equilibrium (see e.g. Chandrasekhar
\cite{chandra}). The simplest description assumes that the halos are
Maclaurin spheroids, which are axisymmetric, oblate, and
homogeneous. Not only is this model fully analytical, it also
provides a convenient background solution to perturb in determining
if and when quantum vortex formation is energetically favored. In
the absence of quantum vortices, however, BEC-CDM is
\textit{irrotational}, while the Maclaurin spheroid model assumes
uniform rotation. Of course, this irrotationality can be broken
locally in the fluid, by creating quantum vortices, if the amount of
angular momentum exceeds a certain minimum, as shown below. In the
limit of large enough angular momentum that a vortex
\textit{lattice} develops, in fact, uniform rotation is a good
approximation, and so will the Maclaurin spheroids be. More
generally, to account for irrotationality, we also consider a second
model, that of irrotational Riemann-S ellipsoids. Since the classic
solution for Riemann-S ellipsoids is homogeneous, however, we
account for the $(n=1)$-polytropic nature of TYPE II BEC-CDM by
adopting the solution derived by Lai, Rasio \& Shapiro (LRS)
\cite{LRS} for \textit{compressible} Riemann-S ellipsoids, based on
the 'ellipsoidal approximation' which assumes self-similar
ellipsoidal density strata.

We denote the semi-axes of those bodies along $(x,y,z)$ as
$(a_1,a_2,a_3)$. Maclaurin spheroids fulfill $a_1=a_2
> a_3$ with eccentricity $e = (1-(a_3/a_1)^2)^{1/2}$. Using
(\ref{virial}), we can determine how the (mean) radius $R =
(a_1a_2a_3)^{1/3}$ and the spin parameter of such a halo depend on
its eccentricity, see RDS \cite{RS}:
  \begin{equation} \label{mac}
   R =
\left(\frac{15}{3A_3(e)(1-e^2)^{2/3}}\right)^{1/2}\left(\frac{g}{4\pi
G m^2}\right)^{1/2},
 \end{equation}
 \begin{equation} \label{lambdaincomp}
 \lambda =
 \frac{6}{5\sqrt{5}}\frac{\arcsin{e}}{e}t\left(1+\frac{e}{t}\frac{A_3(e)(1-e^2)^{1/2}}{\arcsin(e)}\right)^{1/2},
  \end{equation}
 with the $t$-parameter $t \equiv T/|W|$, a measure of rotational support,
  given by (see also LRS \cite{LRS})
 \begin{equation} \label{tpam}
  t(e) = 3/(2e^2) - 1 - 3\sqrt{1-e^2}/(2e\arcsin(e)),
   \end{equation}
    and $A_3(e) = 2/e^2 - 2\sqrt{1-e^2}\arcsin(e)/e^3$ for this
    model.
On the other hand, a compressible, irrotational Riemann-S ellipsoid
of polytropic index $n=1$ must be prolate, i.e. its semi-axes
fulfill $a_1 \geq a_3 \geq a_2$, and the eccentricities are given by
$e_1 = (1-(a_2/a_1)^2)^{1/2}$ and $e_2 = (1-(a_3/a_1)^2)^{1/2}$. In
that case, the expressions for the mean radius and spin parameter
 \begin{equation} \label{rie}
  R = R_0 g(e_1,e_2)^{-1/2},~~\lambda = \lambda(e_1,e_2)
  \end{equation}
   with $R_0$ in (\ref{onesphere}) and $g(e_1,e_2)$ and $\lambda(e_1,e_2)$ functions of the
   eccentricities, are very cumbersome and we refer to RDS \cite{RS}, equ.(101) for
more details.

\begin{figure}
     \begin{minipage}[b]{0.5\linewidth}
      \centering\includegraphics[width=3.7cm]{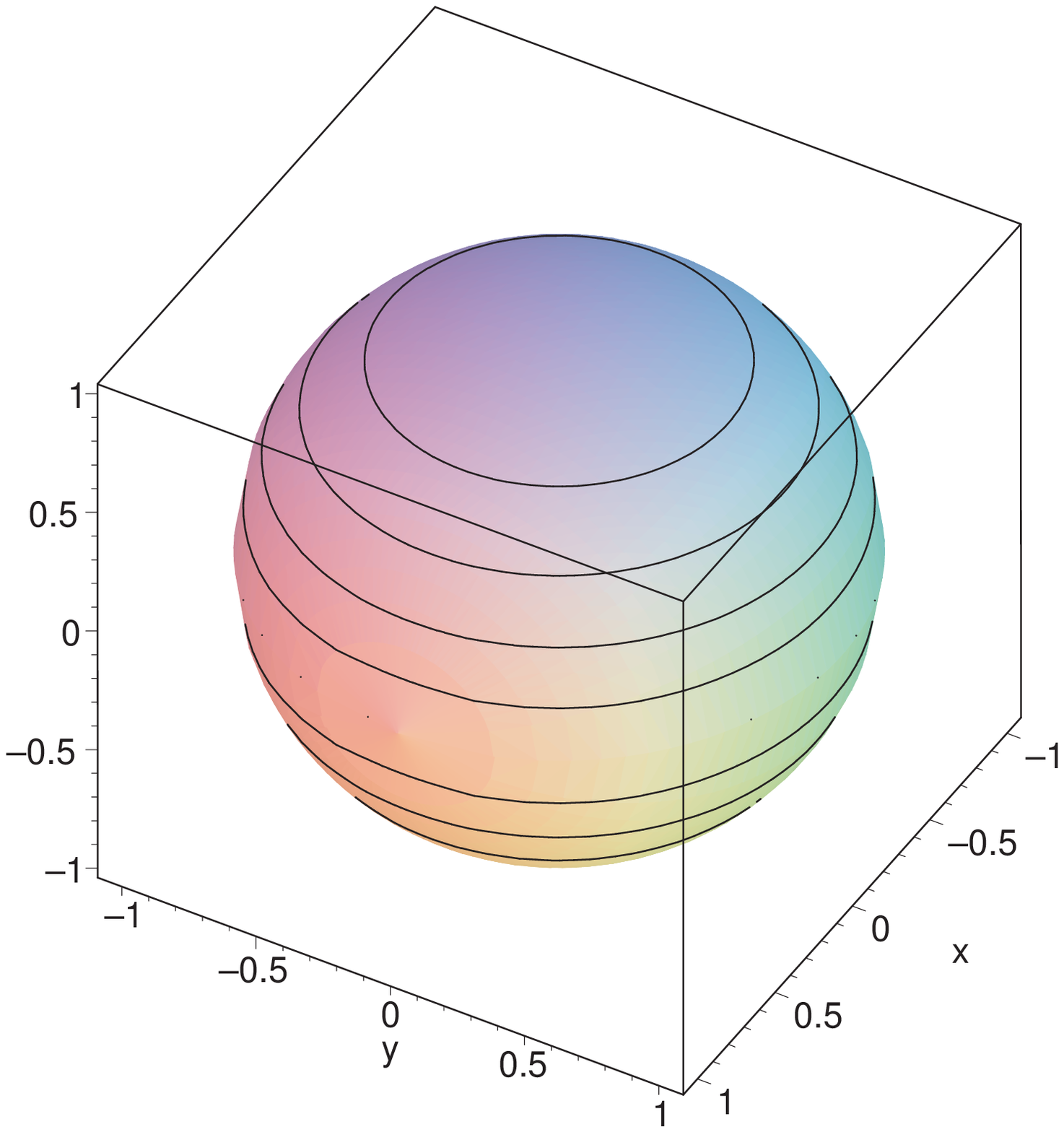}
     \hspace{0.1cm}
    \end{minipage}%
 \begin{minipage}[b]{0.5\linewidth}
      \centering\includegraphics[width=3.7cm]{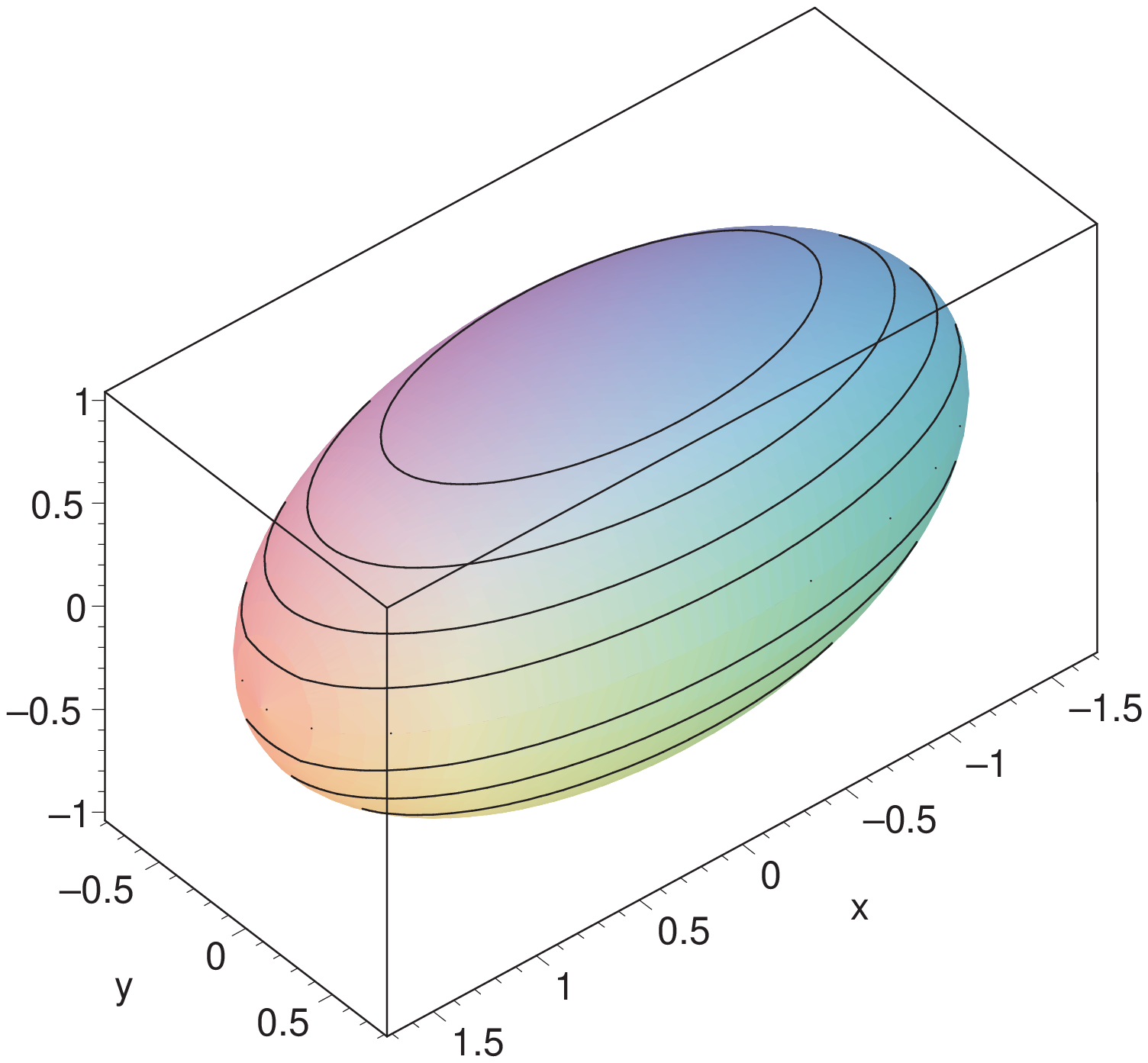}
     \hspace{0.1cm}
    \end{minipage}
 \caption{Halos rotating about the $z$-axis for $a_3 = 1$ and
$\lambda = 0.05$: Maclaurin spheroid having $e = 0.302$
(\textit{left-hand-plot}) and irrotational Riemann-S ellipsoid
having $(e_1,e_2) = (0.881; 0.797)$ (\textit{right-hand-plot}); see
also RDS \cite{RS}.}
 \label{fig:figure1}
\end{figure}

For both models, the characteristic size - the mean radius - depends
on the particle parameters exactly the same way as in the
non-rotating case, namely $R \sim \sqrt{g}/m$. The effect of the
rotation is thus only to change the overall multiplicative factor of
this dependence as seen from (\ref{mac}) and (\ref{rie}). By fixing
$\lambda = (0.01, 0.05, 0.1)$, we can solve $\lambda = \lambda(e)$
in (\ref{lambdaincomp}) and $\lambda = \lambda(e_1,e_2)$ in
(\ref{rie}), respectively, for the eccentricities.
Fig.\ref{fig:figure1} shows two illustrative examples of our
rotating halo models.

\begin{figure}[b]
      \centering\includegraphics[width=7cm]{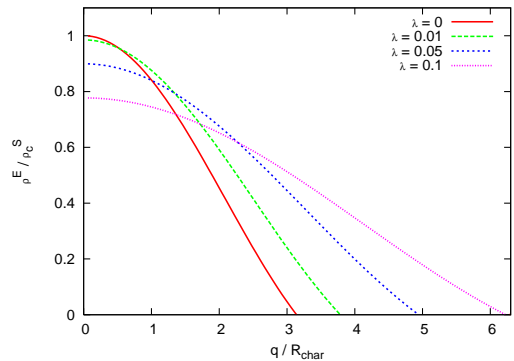}
 \caption{
 Density profiles of the (vortex-free) ($n=1$)-polytropic Riemann-S ellipsoidal halos having
 $\lambda = (0.01, 0.05, 0.1)$, according to equ.(\ref{ellipdensity}).
  The profile of the spherical halo, equ.(\ref{tfprofile}), is added for comparison (solid curve). The densities are
  all normalized to $\rho_c^S = 1$; $q = \tilde q R_{char} \equiv \tilde q [g/(4\pi G m^2)]^{1/2}$. The locus of
  the outer surface, where the density vanishes, increases with $\lambda$; see also RDS \cite{RS}.}
 \label{fig:figure2}
\end{figure}

\begin{figure}[b]
      \centering\includegraphics[width=6cm]{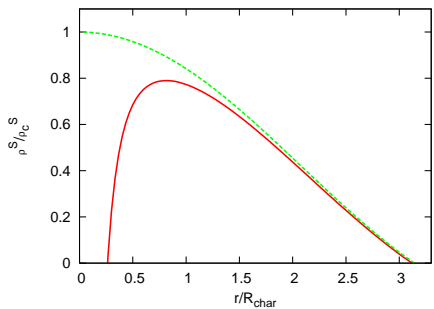}
 \caption{
 Density profile of a spherical halo with vortex in the center (solid line) which transitioned from a Riemann-S ellipsoidal halo having
 $\lambda = 0.05$. The profile of the spherical halo with $L=0$ and no vortex (dashed line) is added for comparison; $R_{char}$ as in Fig. \ref{fig:figure2}.
 See also RDS \cite{RS}.}
 \label{fig:figure3}
\end{figure}

We were able in RDS \cite{RS} to derive the generalization to
equ.(\ref{tfprofile}) for the case of the $(n=1)$-polytropic
Riemann-S ellipsoid analytically, assuming the ellipsoidal
approximation of LRS \cite{LRS}. Accordingly, the ellipsoidal
density profile reads
 \begin{equation} \label{ellipdensity}
     \rho^E (\tilde q) = \rho_c^E sinc \left[\tilde q
     (1-e_1^2)^{1/6}(1-e_2^2)^{1/6}g(e_1,e_2)^{1/2}\right]
     \end{equation}
with
 \begin{displaymath}
  \tilde q^2 = [x^2 + y^2/(1-e_1^2) + z^2/(1-e_2^2)](\pi/R_0)^2,
   \end{displaymath}
  $\rho_c^E$ the central density of the
ellipsoid, and $R_0$ in (\ref{onesphere}). It is plotted for
different $\lambda$ in Fig.\ref{fig:figure2}.

\begin{figure}[b]
      \centering\includegraphics[width=7.2cm]{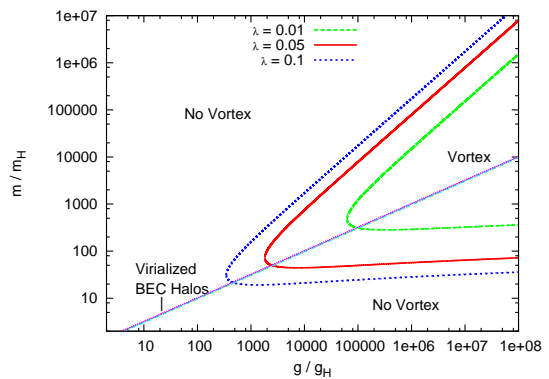}
 \caption{\textit{Model A:} Curves within which vortex formation is energetically favored in the dimensionless BEC-CDM particle parameter space
 $(m/m_H, g/g_H)$ for halos with spin parameter $\lambda = 0.01, 0.05, 0.1$, respectively. Straight lines for the same $\lambda$- values:
 halos fulfilling virial equilibrium, according to equ.(\ref{dmvirial1}); see also RDS \cite{RS}.}
 \label{fig:figure4}
\end{figure}

\begin{figure}[b]
      \centering\includegraphics[width=7.2cm]{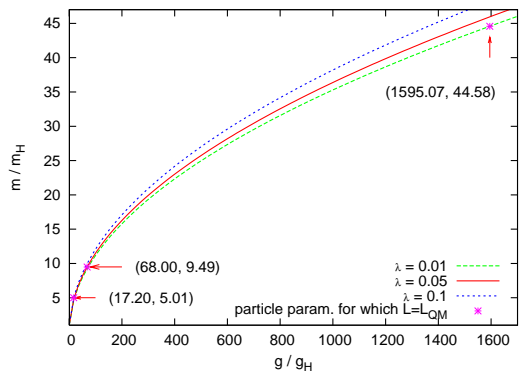}
 \caption{\textit{Model B:} Dimensionless BEC-CDM particle parameter space
 $(m/m_H, g/g_H)$ (no log-scale !) for halos with spin parameter $\lambda = 0.01, 0.05, 0.1$, respectively, fulfilling virial equilibrium,
 according to equ.(\ref{dmvirial2}). For each
 $\lambda$, dots denote those critical particle parameters for which $L = L_{QM}$; see also RDS \cite{RS}.}
 \label{fig:figure5}
\end{figure}

The virial theorem (\ref{virial}) not only provides the above
relationships for the mean radius and spin-parameter of a halo, but
also relates the DM particle parameters in a characteristic way: it
can be shown that the corresponding formulae are
 \begin{equation} \label{dmvirial1}
 \frac{m}{m_H} =
 \left(\frac{5}{8A_3(e)(1-e^2)^{2/3}}\right)^{1/2}\sqrt{\frac{g}{g_H}}
 \end{equation}
 for Maclaurin spheroidal halos, and
\begin{equation} \label{dmvirial2}
 \frac{m}{m_H} =
 \frac{\pi}{\sqrt{8}}g(e_1,e_2)^{-1/2}\sqrt{\frac{g}{g_H}}
 \end{equation}
 for Riemann-S ellipsoidal halos, see RDS \cite{RS}. That is, virialized rotating halos,
 according to either model, will lie
 on a straight line in ($\log m, \log g$)-space, whose slope is
 completely determined by the eccentricities or mean radius of a
 given halo. These relationships can be found in Fig.\ref{fig:figure4} and
 Fig.\ref{fig:figure5}, respectively. From those and the condition $g/g_H \gg
 2$, it follows that, in the TYPE II regime, $m/m_H \gg 1$, as it must be for $\lambda_{deB} \ll R_0$.

In \cite{RS}, we also determined the conditions for the formation of
a central quantum vortex in a given rotating halo. The minimum
amount of angular momentum \textit{necessary} to form a
singly-quantized, axisymmetric vortex in the center of a halo with
total number of particles $N = M/m$ is given by
\begin{equation}
 L_{QM} \equiv N \hbar,
 \end{equation}
which, for a given $\lambda$ and $L$ becomes a minimum condition on
the particle mass. The relationship is linear for both halo models,
i.e.
 \begin{equation} \label{mL}
  m/m_H = f(e_1,e_2)L/L_{QM},
 \end{equation}
 with $f(e_1,e_2)$ denoting here another function depending only on the eccentricities (see equ.(108) and (109) in \cite{RS}).
 In order to find a criterion which is, not only \textit{necessary}, but also \textit{sufficient}, we
 determined the DM particle parameters for which the central vortex is energetically favored, i.e. the conditions
 which make the halo with vortex have less energy than a
 rotating, but otherwise vortex-free halo. For this purpose, we
 consider two limiting cases: \textit{Model A} describes a halo
 with a high enough angular momentum that $L \gg L_{QM}$, i.e. the
 central vortex is essentially considered to be only a small perturbation of
 the total angular momentum $L$ of the halo. For this case, we use Maclaurin spheroids as our halo model.
\textit{Model B}, on the other hand, assumes a given halo has just
enough angular momentum to form one quantum vortex, i.e. $L =
L_{QM}$. We use the irrotational, $(n=1)$-polytropic Riemann-S
ellipsoids for this case. In this case, once the central vortex is
energetically favored, its formation takes up all of the angular
momentum, leaving a spherical halo with vortex, see
Fig.\ref{fig:figure3} for an example.

Despite differences in the assumed halo equilibrium models, the
conclusions are the same for \textit{Model A} and \textit{Model B}:
vortex formation requires a minimum particle mass $m \geq m_{crit}$
and a minimum particle self-interaction coupling strength $g \geq
g_{crit}$. For any $(m,g)$ pair which satisfies $m \geq m_{crit}$
and $g \geq g_{crit}$ at a given $\lambda$, this pair will also
favor vortex formation for any \textit{larger} value of $\lambda$.
Our main results are summarized in Fig.\ref{fig:figure4} and
\ref{fig:figure5}, which are independent of halo size, respectively,
while Table \ref{tab:table2} and \ref{tab:table3} list physical
units for $m_{crit}, g_{crit}$ and the corresponding vortex size,
for the cases of a typical dwarf galaxy (DG) and a dwarf spheroidal
galaxy (dSph), according to Table \ref{tab:table1}. Since we fix
$\lambda$ and $L = L_{QM}$ for the case of \textit{Model B}, the
respective particle parameters $((m/m_H)_{crit}, (g/g_H)_{crit})$
are uniquely determined via equ.(\ref{mL}) and (\ref{dmvirial2}),
respectively (see also Fig. \ref{fig:figure5} for those numbers). It
turns out that for all $\lambda$ considered, vortex formation is
favored for those parameters. We interpret this to mean that, for
these same $\lambda$-values, if $L/L_{QM} > 1$, instead, (i.e.
$m/m_H > (m/m_H)_{crit}$, according to (\ref{mL})), vortex formation
will also be favored, as long as $g/g_H > (g/g_H)_{crit}$. On the
other hand, for halos with $L < L_{QM}$, the particle parameters
must satisfy $m/m_H < (m/m_H)_{crit}$ and $g/g_H < (g/g_H)_{crit}$,
respectively, so they will \textit{not} form vortices, but can still
be modeled as Riemann-S ellipsoids. (This assumes, of course, that
$g/g_H \gg 2$, so they are still in the TYPE II regime.)

We shall comment on other previous work on vortices in related
models. Silverman \& Mallet \cite{SM} give heuristic arguments for
vortex formation, but do not derive the critical conditions or their
consequences for the DM particle parameters. Yu \& Morgan \cite{YM}
show that vortex lattices can provide flat rotation curves for
galaxies. Kain and Ling \cite{kain}, on the other hand, find
approximate solutions for the density profile of a nonrotating,
spherically-symmetric halo with a single vortex that contains all
the angular momentum for the TYPE II case (as in Fig.
\ref{fig:figure3}, solid curve). Their estimates for the viable
parameter space of particle mass for the Andromeda galaxy is in
agreement with our more precise results for the Milky Way, see RDS
\cite{RS}.

\begin{table}[tb]
\caption{\label{tab:table2}
 Lower bounds on the boson mass and self-interaction coupling strength for vortex formation in TYPE II BEC-CDM halos with a given spin
 parameter $\lambda$ in \textit{Model A}. $\xi_{max}$-values denote the upper bounds for the corresponding vortex core radius.}
\begin{ruledtabular}
Dwarf Galaxy
\begin{tabular}{llll}
 $\lambda$ & $m_{crit}$
\textrm{[eV]} & $g_{crit}$ \textrm{[eV cm$^3$]}& $\xi_{max}$ \textrm{[kpc]}\\
 \hline
 \colrule
$0.01$ & $1.04 \cdot 10^{-21}$ & $2.30\cdot 10^{-58}$ & $0.03$\\
$0.05$ & $1.67 \cdot 10^{-22}$ & $5.74\cdot 10^{-60}$ & $0.20$\\
$0.10$ & $7.33 \cdot 10^{-23}$ & $1.02\cdot 10^{-60}$ & $0.47$\\
 \end{tabular}

Dwarf Spheroidal Galaxy
 \begin{tabular}{llll}
 $\lambda$ & $m_{crit}$
\textrm{[eV]} & $g_{crit}$ \textrm{[eV cm$^3$]}& $\xi_{max}$ \textrm{[kpc]}\\
 \hline
 \colrule
$0.01$ & $3.30 \cdot 10^{-20}$ & $2.30\cdot 10^{-57}$ & $3.13\cdot 10^{-3}$\\
$0.05$ & $5.28 \cdot 10^{-21}$ & $5.74\cdot 10^{-59}$ & $0.02$\\
$0.10$ & $2.32 \cdot 10^{-21}$ & $1.02\cdot 10^{-59}$ & $0.05$\\
\end{tabular}
\end{ruledtabular}
\end{table}

\begin{table}[tb]
\caption{\label{tab:table3}
 Lower bounds on the boson mass and self-interaction coupling strength for vortex formation in TYPE II BEC-CDM halos with a given spin
 parameter $\lambda$ in \textit{Model B}. $\xi_{max}$-values denote the upper bounds for the corresponding vortex core radius.}
\begin{ruledtabular}
Dwarf Galaxy
\begin{tabular}{llll}
 $\lambda$ & $m_{crit}$
\textrm{[eV]} & $g_{crit}$ \textrm{[eV cm$^3$]}& $\xi_{max}$ \textrm{[kpc]}\\
 \hline
 \colrule
$0.01$ & $1.50 \cdot 10^{-22}$ & $3.59\cdot 10^{-60}$ & $0.25$\\
$0.05$ & $3.20 \cdot 10^{-23}$ & $1.53\cdot 10^{-61}$ & $1.21$\\
$0.10$ & $1.69 \cdot 10^{-23}$ & $3.87\cdot 10^{-62}$ & $2.41$\\
 \end{tabular}
Dwarf Spheroidal Galaxy
 \begin{tabular}{llll}
 $\lambda$ & $m_{crit}$
\textrm{[eV]} & $g_{crit}$ \textrm{[eV cm$^3$]}& $\xi_{max}$ \textrm{[kpc]}\\
 \hline
 \colrule
$0.01$ & $4.75 \cdot 10^{-21}$ & $3.59\cdot 10^{-59}$ & $0.02$\\
$0.05$ & $1.01 \cdot 10^{-21}$ & $1.53\cdot 10^{-60}$ & $0.12$\\
$0.10$ & $5.34 \cdot 10^{-22}$ & $3.87\cdot 10^{-61}$ & $0.24$\\
\end{tabular}
\end{ruledtabular}
\end{table}

Our results strongly suggest that vortices can only form in BEC-CDM
with a sufficiently large, positive self-interaction coupling
strength. Vortex formation in axion DM \textit{without
self-interaction}, which has been claimed by Sikivie \& Yang
\cite{SY}, seems thus not viable according to our results. However,
for TYPE II halos, i.e. those for which $m/m_H \gg 1$ and $g/g_H \gg
2$, vortex formation is favored for the range of halo spin
parameters found in the CDM model, for a large portion of the BEC
particle parameter space. For example, for $\lambda = 0.05$,
\textit{Model B} yields $(m/m_H)_{crit} = 9.49$ and $(g/g_H)_{crit}
= 68.00$, while \textit{Model A} yields $(m/m_H)_{crit} = 49.52$ and
$(g/g_H)_{crit} = 2549.24$ (see Fig. \ref{fig:figure4} and
\ref{fig:figure5}). Since the presence of a vortex causes the DM
density profile to drop inside the vortex core, diminishing it
altogether at the very center, the effect of vortices must be
seriously considered whenever the TYPE II regime is studied as a
model to describe galactic halo dynamics. At the critical values for
vortex formation and a little above, the effect of the vortex could
be observable, as the $\xi_{max}$-values in Table \ref{tab:table2}
and \ref{tab:table3} suggest. However, while it is true that the
vortex becomes increasingly favored for large $g$ (at fixed
$\lambda$), its size relative to the halo size decreases, i.e. for
large enough $g$ the influence of the vortex will yet again
diminish.

\subsection{\label{sec:leve34}Influence on baryonic substructures}

The foregoing sections exemplified some distinctive characteristics
of Bose-Einstein-condensed DM models as compared to standard
collisionless CDM. A further interesting topic is the detailed
dynamics of baryonic substructure within galactic halos made of
BEC-CDM. While this may likewise provide evidence for the presence
or absence of this form of DM, it has hardly been considered in the
previous literature.

Goodman \cite{goodman} seems to be the first to discuss the
consequences of TYPE II, as a superfluid, for a rotating galactic
bar. This discussion has been continued in Appendix B of Slepian \&
Goodman \cite{SG}, with the result that the drag on the bar can be
smaller than estimated for standard CDM, providing a potential
remedy to a problem which plagues the latter.

Lora at el. \cite{lora}, on the other hand, have recently studied
the dynamic survival of cold gas clumps and globular clusters in the
dSph galaxies of Ursa Minor and Fornax, respectively, in both TYPE I
and TYPE II regimes. The survival of those structures within the
lifetime of these galaxies of known mass provides valuable
constraints on the allowed boson mass and self-interaction coupling
strength. For TYPE I, their calculations favor a boson mass between
$0.3 \cdot 10^{-22} ~\rm{eV} < m < 10^{-22} ~\rm{eV}$, while $m$ can
be larger for TYPE II at fixed coupling strength, their constraint
reading $g/m^2 \gtrsim 8 \cdot 10^{-19}$ cm$^3$/eV (expressed in our
units), corresponding to core sizes larger than about $R \gtrsim
0.7$ kpc. The first result can be understood given the limits in
Table \ref{tab:table1}, while the second result is in accordance
with previous studies, including our own ones as presented in Section
\ref{sec:level2}. In both cases, Lora et al.\cite{lora} find viable
parts of the parameter space of BEC-CDM which can explain the
observations, while others can be definitely excluded.

The study of the dynamics of baryonic 'test bodies' like globular
clusters or giant molecular clouds, which are large enough to feel
the influence of the subtleties in the DM distribution, while still
being much smaller than the DM halo, will provide useful constraints
on BEC-CDM for our own Galaxy. There is valuable and interesting
work to be done in future studies of this problem.

Finally, we note that galaxies are often observed to harbor
supermassive black holes at their centers, with associated quasar
luminosity which supports the idea that the black hole mass grew
primarily by baryonic accretion. One might ask if BEC-CDM is
consistent with this phenomenon, since its fluid behavior might
imply a much higher accretion rate than that of collisionless CDM
particles (c.f. Shapiro \& Teukolsky \cite{ST}). The gravitational
collapse of non-self-interacting scalar field dark matter onto a
central Schwarzschild black hole in a spherically-symmetric
space-time has been considered, for example, by Barranco et
al.\cite{barranco} and references therein, to determine if it is
possible for DM halos in this model to survive for cosmological time
scales. As we discussed in Section \ref{sec:leve32}, however, halos
have angular momentum. For accretion onto a central black hole to
occur, this angular momentum must be overcome, which usually depends
upon a dissipative process involving some form of viscosity to
transfer that angular momentum outward. As a frictionless
superfluid, BEC-CDM halos, however, cannot do this. Moreover, if a
vortex is formed at the center, the lower density there will further
inhibit accretion.

\section{\label{sec:level4}Bounds on particle mass and coupling strength}

We have seen that the equilibrium size of isolated, hydrostatic
BEC-CDM halo structures is predominantly governed either by quantum
pressure, equ.(\ref{qpot}), or by the polytropic pressure due to the
collective, repulsive self-interaction of the DM particles,
equ.(\ref{selfpressure}), for $g \lesssim g_H$ or $g \gtrsim g_H$,
respectively. By opposing gravity, quantum and self-interaction
pressures each prevent structure from forming on small scales,
thereby imposing lower limits on the size of structures we can
expect to find in a BEC-CDM universe. The lower limit set by quantum
pressure is $\lambda_{deB}$ in (\ref{deB}), evaluated using the halo
virial velocity (which also characterizes the bulk mass motions at
its formation time). Since the smallest halos also have the smallest
virial velocities, the most stringent limit results if we require
$\lambda_{deB} \lesssim R$ for the smallest halos. This, in turn,
imposes the lower limit on $m$, see Table \ref{tab:table1}. For a
given $m$ that satisfies this lower limit, the coupling strength $g$
must not exceed the value such that the radius of the polytrope
supported by self-interaction pressure, $R_0$ in (\ref{onesphere})
or the respective generalizations for rotating halos in (\ref{mac})
and (\ref{rie}), exceeds the size of the smallest-scale structure.

The same requirement that the characteristic size $R_0$ not exceed
the size $R$ of the smallest halos can also be used to place an
\textit{upper} limit on particle mass $m$, if we can establish an
upper bound on the 2-body scattering cross section per unit particle
mass, $\sigma/m$, from some other argument. Slepian and Goodman
\cite{SG} have argued that upper bounds on $\sigma/m$ for the
elastic-scattering particles in the SIDM model, based upon comparing
that model to astronomical observation, should apply to BEC-CDM, as
well. The interpretation of the Bullet cluster observations, for
example, as a nearly collisionless merger of two cluster-sized halos
has been found to limit $\sigma/m$ for SIDM halos to
$(\sigma/m)_{max} < 1.25$ cm$^2$/g, according to Randall et
al.\cite{randall}. We can relate $\sigma/m$ for TYPE II BEC-CDM to
the characteristic size $R_0$ in equ.(\ref{onesphere}) according to
\begin{equation} \label{cross}
\frac{\sigma}{m} = \frac{8 G^2}{\pi^3 \hbar^4}R_0^4 m^5,
\end{equation}
as pointed out by Slepian and Goodman \cite{SG}. In order for
(\ref{cross}) to be applicable, the particle mass must be such that
$m/m_H \gg 1$, in which case RDS \cite{RS} noted that (\ref{cross})
can be rewritten in fiducial units related to the size $R$ and mass
$M$ of a given halo or halo core, as follows:
 \begin{displaymath}
  \frac{\sigma}{m} =
   2.094\cdot
  10^{-95}\left(\frac{m}{m_H}\right)^5 \times
   \end{displaymath}
   \begin{equation} \label{sigmaTF}
   \times \left(\frac{R}{1~\rm{kpc}}\right)^{3/2}
  \left(\frac{M}{10^{8}~M_{\odot}}\right)^{-5/2}~\frac{\rm{cm}^2}{\rm{g}},
  \end{equation}
  (where 'g' here means 'grams', not coupling strength). This shows
  that $\sigma/m$ is very much smaller than $(\sigma/m)_{max} \sim
  1$ cm$^2$/g, unless the particle mass is many orders of magnitude
  larger than the lower bounds, $m/m_H = 1$, in Table
  \ref{tab:table1}. We can place an \textit{upper} bound on $m$, in
  fact, if we replace $R_0$ in equ.(\ref{cross}) by $R$, the
  smallest halo (or halo core) size which must be produced, and
  replace $\sigma/m$ by $(\sigma/m)_{max}$, to write
 \begin{equation}
  m < \left(\frac{\pi^3 \hbar^4}{8G^2}\right)^{1/5}\left(\frac{\sigma}{m}\right)_{max}^{1/5}R^{-4/5}.
   \end{equation}
If we take as our fiducial units $R = 1$ kpc, and
$(\sigma/m)_{max}=1$ cm$^2$/g, this gives
  \begin{equation} \label{mupper}
   m < 9.193 \cdot 10^{-4} ~ \rm{eV}/c^2.
   \end{equation}
To be self-consistent, we must check if our assumption is valid that
BEC-CDM remains a pure condensate without thermalizing during
virialization. If the relaxation time for particle collisions to
establish thermodynamic equilibrium at the halo virial temperature
is less than a Hubble time ($\sim 10^{17}$ sec), our assumption
would break down. In that case, Slepian and Goodman \cite{SG}
determined that BEC-CDM halos in the TYPE II regime would have cores
surrounded by isothermal envelopes of non-condensate, which would
yield density profiles in disagreement with observed halos (unless
$\sigma/m$ exceeds $(\sigma/m)_{max} \sim 1$ cm$^2$/g by orders of
magnitude). It can be shown, however, that if $(\sigma/m)_{max} \sim
1$ cm$^2$/g, then the relaxation time for achieving thermodynamic
equilibrium is, indeed, more than a Hubble time, so thermodynamic
equilibrium is \textit{not} achieved for particle masses which obey
inequality (\ref{mupper}). Hence, this upper limit is a
self-consistent one. Apparently, there is quite a large range of
particle mass allowed between these upper and lower limits.

\section{\label{sec:level5}Beyond the polytropic size limit}

Nevertheless, this requirement that $R_0 < R$ for the smallest halos
suggests there is a problem for this model in its simplest form, if
observations require us to accommodate the formation of objects as
small as the smallest dwarf spheroidal galaxies, while at the same
time serving to explain the flattening of the density profiles in
the cores of much larger galaxies. The remedy suggested by Slepian
\& Goodman \cite{SG}, where BEC-CDM cores in the TYPE II regime are
enshrouded by isothermal envelopes of non-condensate, was
unfortunately shown not to work by the same authors.

We envisage a different scenario to overcome the size limit given by
the equilibrium model as follows: we assume that (\ref{onesphere})
characterizes the size of the inner core of larger halos, which grow
larger as a result of continuous infall at the time of halo
formation. Let us sketch this here in more detail. It can be shown
that a classic, spherical top-hat model for the collapse and
virialization of a cosmological density perturbation applied to
BEC-CDM in the Einstein-de Sitter universe yields a post-collapse
virialized object with radius $R_{TH}$ and uniform density $\rho_0$,
given by
 \begin{equation} \label{thratio}
  \frac{R_{TH}}{r_{ta}} = \frac{2}{3} \mbox{  and  }
 \frac{\rho_0}{\rho_{ta}} = \left(\frac{3}{2}\right)^3,
  \end{equation}
  where $r_{ta}$ and $\rho_{ta}$ are the radius and density of the top-hat at the time of turn-around (i.e. maximum
  expansion). Then, the (virial)
  radius is determined from (\ref{virial}) as
  \begin{equation} \label{zerosphere}
  R_{TH} = \sqrt{\frac{15}{2}}\left(\frac{g}{4\pi
  G m^2}\right)^{1/2}.
   \end{equation}
Not surprisingly, this $R_{TH}$ depends upon $g$ and $m$ in the same
way the polytrope radius $R_0$ does, so the implication is that
halos of different mass must form from cosmological fluctuations
that collapse at different times, in order that $\rho_{ta}$ is
different. Unfortunately, more massive halos require higher
$\rho_{ta}$, since $R_{TH}$ is independent of halo mass, but higher
$\rho_{ta}$ requires collapse at earlier times, which reverses the
hierarchical structure formation history expected for CDM.

The crucial idea in overcoming this undesirable feature is the
realization that equ.(\ref{zerosphere}) neglects any internal
kinetic energy of the virialized object. In fact, it can be shown
that the virial radius grows beyond $R_{TH}$, once an (effective)
kinetic term has been added, even though the regime of TYPE II is
retained. Determining the form and the physical meaning of this
additional kinetic part, and how it can advance the above
description to provide a successful model for halo formation and
structure will be the subject of a forthcoming paper. An immediate
observation is the fact that the effective kinetic energy $K_{eff}$
should not simply be a constant factor times $W$ or $U_{SI}$, since
this will only increase the virial radius by a fixed factor times
$R_{TH}$, as can be easily shown. In fact, we have seen that the
inclusion of uniform rotation has had exactly this effect, see
equ.(\ref{mac}, \ref{rie}). Instead, we will pursue the following
idea: the BEC-CDM fluid must undergo oscillations during the process
of virialization due to its inherently quantum-mechanical nature. In
fact, the results of work by Khlopov et al.\cite{khlopov} and
G\'uzman \& Ure\~na-L\'opez \cite{GU} lend support to this idea.

Suppose $K_{eff}$ describes wave motions. For the general argument
outlined here, it is sufficient to consider a non-vanishing bulk
velocity $\mathbf{v}$ as before, i.e. $K_{eff} = T = \int
\frac{\rho_0}{2} \mathbf{v}^2 dV
> 0$.
Since $\rho_0 = const.$, we assume that the gross average in the
form of $T = \frac{3}{2} M \sigma_v^2$ with velocity dispersion
$\sigma_v^2 = \frac{1}{3}\langle \mathbf{v}^2\rangle$, will capture
the overall kinetic contribution due to wave motions. Additionally,
in order for the virial radius to depend on halo mass and time of
collapse $t_{coll}$, as they do for standard CDM, we require the
top-hat density to be a fixed fraction of the background density at
$t_{coll}$, that is, $\rho_0 = A \rho_{b, coll}$, with the constant
$A$ depending on the background cosmology, e.g. $A \simeq 178$ for
standard CDM in a flat, matter-dominated universe. The
  corresponding total mass is then $M = \frac{4}{3}\pi R^3 A
  \rho_{b, coll}$, which yields a radius
   \begin{equation} \label{radius}
  R(M,t_{coll}) = \left(\frac{3M}{4\pi
  A}\frac{1}{\rho_{b, coll}}\right)^{1/3}.
  \end{equation}
 Inserting this expression for $R$ into (\ref{virial}) with $K_Q = 0$ and solving
 for the unknown velocity dispersion, we get
  \begin{equation} \label{vd}
  \sigma_v^2 (M,t_{coll}) = \frac{(36\pi)^{1/3}}{15} G M^{2/3}
  (A\rho_{b, coll})^{1/3} - \frac{g}{2m^2} A \rho_{b, coll}.
  \end{equation}
To interpret this result, we observe the following: For $g=0$, the
last term vanishes and we recover the standard case, but for
increasing $g > 0$, this term makes $\sigma_v$ smaller. We can
calculate the minimum mass for which $\sigma_v = 0$, resulting in
 \begin{displaymath}
 M_{min} =
 \left(\frac{15}{(36\pi)^{1/3}G}\right)^{3/2}A\left(\frac{g}{2m^2}\right)^{3/2}\rho_{b,
 coll} =
  \end{displaymath}
   \begin{equation}
   = \frac{4}{3}\pi A \rho_{b,coll} R_{TH,0}^3
 \end{equation}
  with $R_{TH,0}$ given by (\ref{zerosphere}). Thus, the
  smallest mass halo has the minimum size of $R_{TH,0}$ by construction. The particle
  parameters may now be chosen such that $M_{min}$ corresponds to
  the smallest observed galaxies,
  as well as to the cores of large galaxies, e.g. $M_{min} \simeq 10^8 ~M_{\odot}$.
   Larger halos then follow the relationship (\ref{radius}) and have a non-vanishing
  velocity dispersion due to internal wave motion, according to
  (\ref{vd}). This guarantees that BEC-CDM halos of mass $M >
  M_{min}$ would share the mass-radius relation of halos in the
  standard CDM model, if halos of a given mass $M$ typically
  collapse at the same time as they do for standard CDM.

\begin{acknowledgments}
TRD would like to thank Tonatiuh Matos and Claudia Moreno and the
organizing committee for their kind hospitality at the \textit{IV
International Meeting on Gravitation and Cosmology}, Guadalajara,
Mexico, May 21-25, 2012. This work was supported in part by U.S. NSF
grants AST-0708176, AST-1009799 and NASA grants NNX07AH09G,
NNG04G177G, NNX11AE09G to PRS. TRD also acknowledges support by the
Texas Cosmology Center of the University of Texas at Austin.
\end{acknowledgments}

\bibliography{apssamp}

\end{document}